\documentclass{webofc}
\usepackage[varg]{txfonts}   % Web of Conferences font
\usepackage{hyperref}
\usepackage{url}
\usepackage{graphicx}
\usepackage{color}
\usepackage[dvipsnames]{xcolor}

\hypersetup{colorlinks=true,citecolor=blue,urlcolor=blue,linkcolor=blue}
\begin{document}
\title{Electromagnetic radiation from Quark-Gluon Plasma at finite baryon density}

\author{
\firstname{Xiang-Yu} \lastname{Wu}\inst{1}\fnsep\thanks{Speaker, \email{xiangyu.wu2@mail.mcgill.ca}} 
\and
\firstname{Charles} \lastname{Gale}\inst{1}\fnsep\thanks{\email{charles.gale@mcgill.ca}} 
\and
\firstname{Sangyong} \lastname{Jeon}\inst{1}\fnsep\thanks{\email{sangyong.jeon@mcgill.ca}}
\and
\firstname{Jean-Fran\c{c}ois} \lastname{Paquet}\inst{2}\fnsep\thanks{\email{jean-francois.paquet@vanderbilt.edu}}
\and
\firstname{Bj\"orn } \lastname{Schenke}\inst{3}\fnsep\thanks{\email{bschenke@bnl.gov}}
\and
\firstname{Chun } \lastname{Shen}\inst{4}\fnsep\thanks{\email{chunshen@wayne.edu}}
        % etc.
}

\institute{Department of Physics, McGill University, 3600 University Street, Montreal, Qu\'ebec H3A 2T8, Canada
\and
           Department of Physics and Astronomy, Vanderbilt University, Nashville, Tennessee 37240, USA
\and
           Physics Department, Brookhaven National Laboratory, Upton, New York 11973, USA
\and
            Department of Physics and Astronomy, Wayne State University, Detroit, Michigan 48201, USA
          }

% \institute{Insert the first address here 
% \and
%            the second here 
% \and
%            Last address
%           }

\abstract{ Using the Bayesian calibrated iEBE-MUSIC framework, we compute the production of electromagnetic radiation from hot hadronic matter at finite baryon density. Results for thermal photon and thermal dilepton yields are obtained by folding in-medium emission rates with posterior-sampled backgrounds evolved hydrodynamically.
We consider different photon sources and analyze the collision-energy dependence of the thermal-to-prompt photon ratio. 
The sensitivity of the dilepton spectra to the pre-equilibrium stage  is explored by considering different initialization procedures. 
Finally, we examine the impact on dilepton spectra of choosing parameter sets stemming from different Bayesian data analyses.
}
\maketitle
\section{Introduction}
\label{intro}
Experiments involving the collisions of relativistic heavy-ions performed at RHIC (Relativistic Heavy Ion Collider) in the Beam Energy Scan (BES) program  and at the CERN Super Proton Synchrotron (SPS)  provide the   opportunity to explore the QCD phase diagram at temperatures near the QCD crossover transition temperature, but at finite baryon density. 
In recent years, multistage approaches with a hydrodynamic core have become a standard modeling tool in the field, and are widely used to quantify the properties of the quark-gluon plasma (QGP) via hadronic observables. 
As an essential complement to hadronic signals, electromagnetic (EM) observables such as photons and dileptons constitute valuable and direct observables for investigating the properties of the QGP medium, as they are produced throughout all stages of the evolution of the system and escape the strongly interacting environment  with minimal final-state interactions.
With appropriately chosen kinematic cuts for photons and dileptons, they can reveal information about the early stages of heavy ion collisions.

\section{Theoretical Framework}
\label{sec-1}
The spacetime evolution of relativistic heavy-ion collisions at finite baryon density is simulated using the  iEBE-MUSIC framework, combining 3D dynamical initial conditions~\cite{Shen:2022oyg} with viscous fluid dynamics (MUSIC~\cite{Paquet:2015lta, Denicol:2018wdp})  and hadronic afterburner (UrQMD).
We use Bayesian inference with hadronic data from the RHIC-BES to calibrate the model parameters ~\cite{Jahan:2024wpj}. After model calibration, thermal photon and thermal dilepton production can be calculated by convolving the thermal emission rates with local fluid cell information, including space-time volume $V$, local temperature $T$, local flow velocity $u^{\mu}$, and local baryon chemical potential $\mu_B$. In the case of photons, both thermal partonic photon emission rates and thermal hadronic photon emission rates, each including viscosity corrections, are considered~\cite{Paquet:2015lta}. For dilepton production, only thermal dileptons from the QGP phase~\cite{Churchill:2023vpt} are considered here. The EM rates switch from QCD-based to hadronic-based at a temperature $T = 160$ MeV, and the hadronic phase ends at T = 120 MeV.  Finally, in order to obtain the direct photon production, we also include the prompt photon contribution calculated with INCNLO~\cite{Aurenche:1998gv}, which is based on NLO pQCD (perturbative QCD evaluated at next-to-leading order in the coupling).

\section{Results}
\label{sec-2}

\begin{figure}[h]
\centering
\includegraphics[width=5cm,clip]{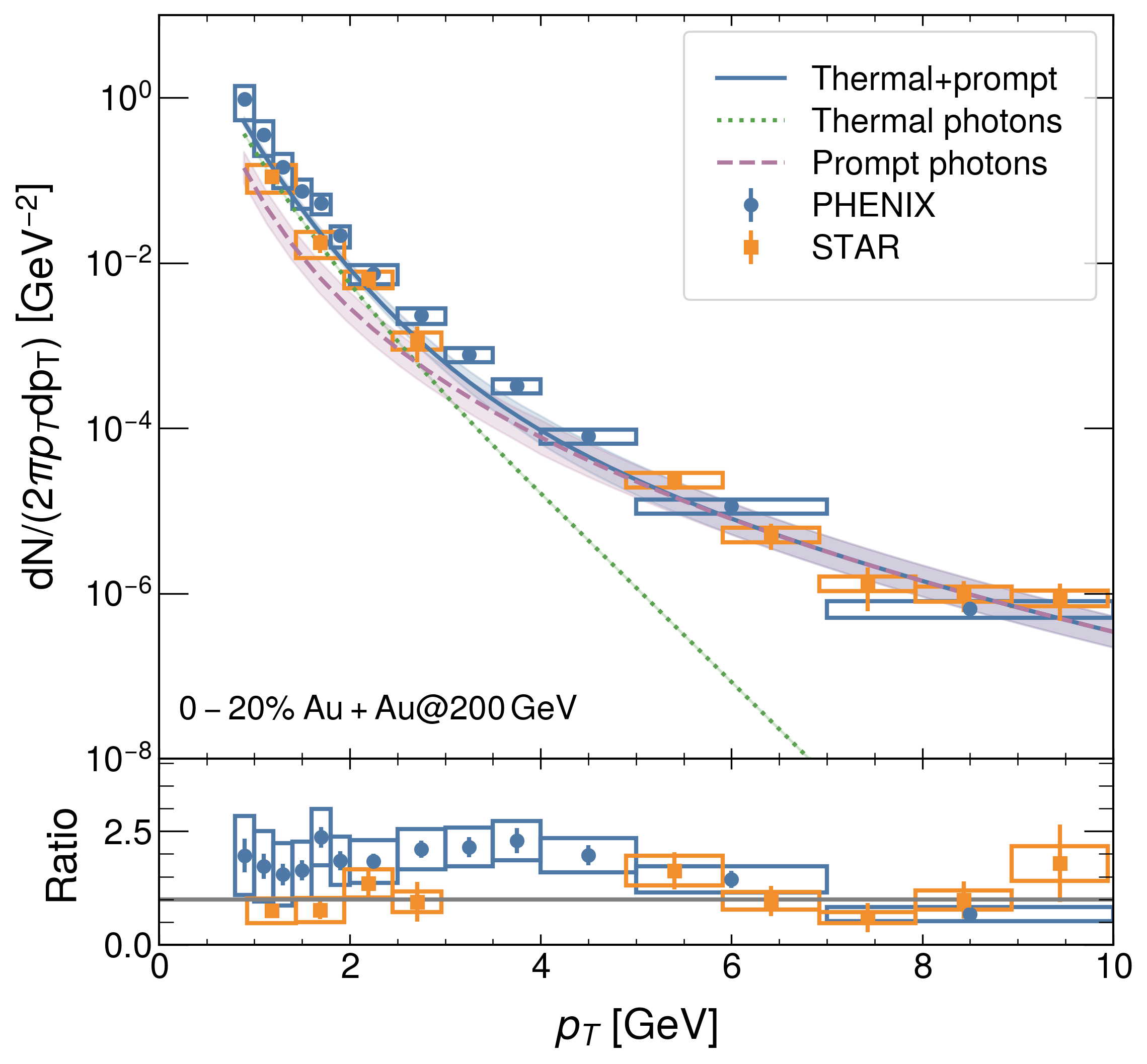}
\includegraphics[width=6cm,clip]{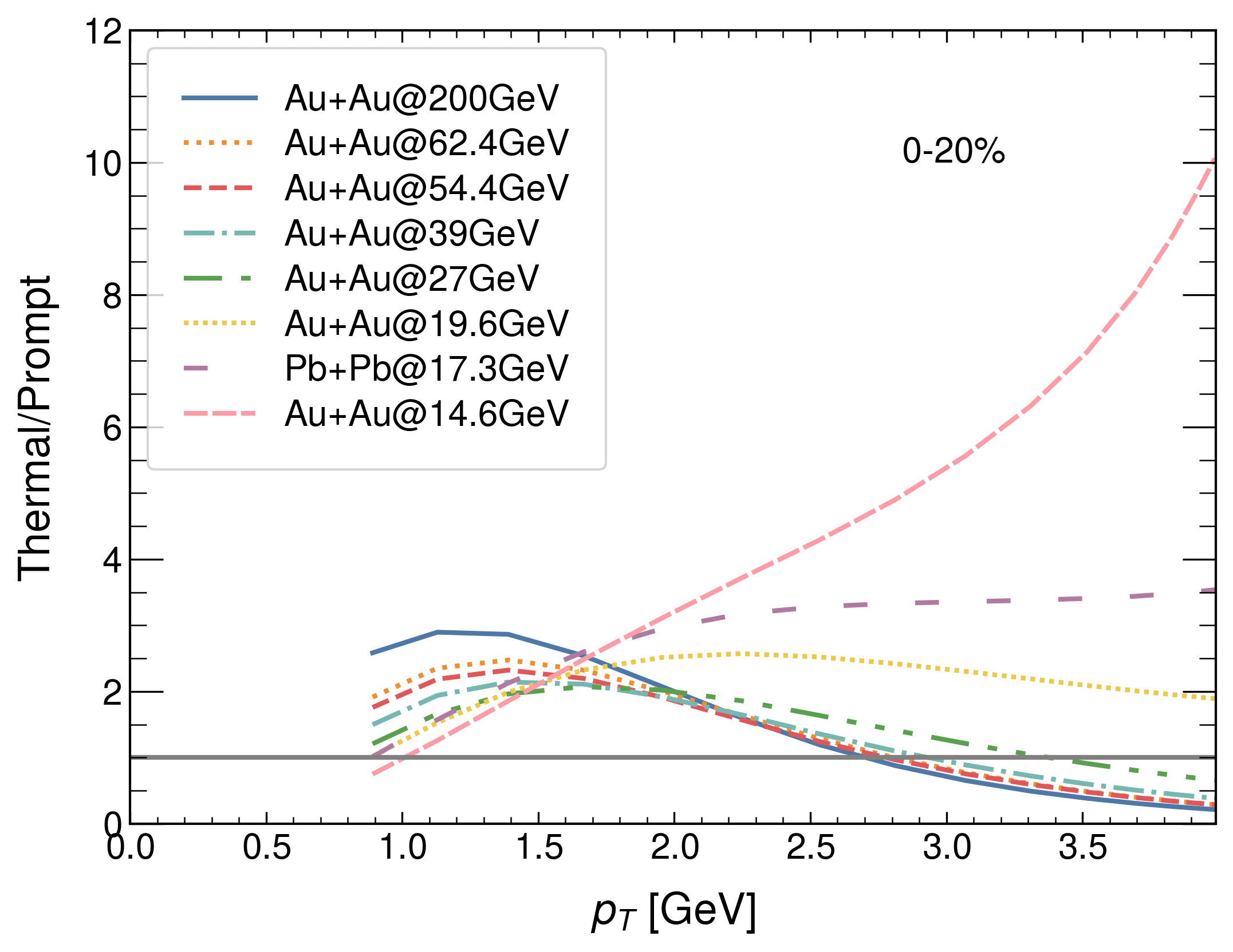}
\caption{(Left panel): Direct photon yield as a function of transverse momentum, for 0–20\% centrality class Au+Au collisions at 200 GeV. The total yield and its prompt and thermal components are shown. Data are from STAR~\cite{STAR:2016use} and PHENIX~\cite{PHENIX:2014nkk} Collaborations. The lower subpanel shows the ratio between data and the calculated total yield. (Right panel): Ratio of thermal to prompt photon yields as a function of transverse momentum for Au+Au collisions in the 0–20\% centrality class at $\sqrt{s_{\rm NN}}=$14.6–200 GeV, and for Pb+Pb collisions at $\sqrt{s_{\rm NN}}=$ 17.3 GeV.}
\label{fig-1}       % Give a unique label
\end{figure}
The left panel of Fig.~\ref{fig-1} shows the direct photon yield and its prompt and thermal component for 0–20\% central Au+Au collisions, with the data to calculated direct photon ratio in the lower subpanel. Here, the uncertainty in the prompt photon calculation is estimated by varying the renormalization and factorization scales from 0.5$p_T$ to 2.0$p_T$. The calculated prompt photons dominate at high $p_T$ and agree with STAR~\cite{STAR:2016use} and PHENIX~\cite{PHENIX:2014nkk} data within error bars. This indicates that $N_{\rm coll}$-scaling works well in that region. For low $p_T$ ($p_T\lesssim2.5$~GeV), thermal emission dominates and the prompt component is subleading. Overall, our direct photon result is consistent with STAR, whereas it underestimates the PHENIX measurement in this low $p_T$ region. 

The right panel of Fig.\ref{fig-1} shows the ratio of thermal to prompt photon yield as a function of $p_T$ in the 0–20\% centrality class of Au+Au collisions at energies $\sqrt{s_{NN}}$ = 14.6–200~GeV and Pb+Pb collisions at 17.3 GeV. 
In the range of $p_T \lesssim 2.5$ GeV, thermal photons dominate over prompt photons at all collision energies. For $p_T \gtrsim 2.5$ GeV, prompt photons dominate at high collision energies, as also seen in the left panel of Fig.\ref{fig-1}. However, 
as the collision energy decreases, our calculations of prompt photons fall steeply at large Bjorken-$x$.
As a result, thermal photons begin to dominate over prompt photons even in the high $p_T$ region, making their identification less ambiguous. 

\begin{figure}[h]
\centering
\includegraphics[width=6cm,clip]{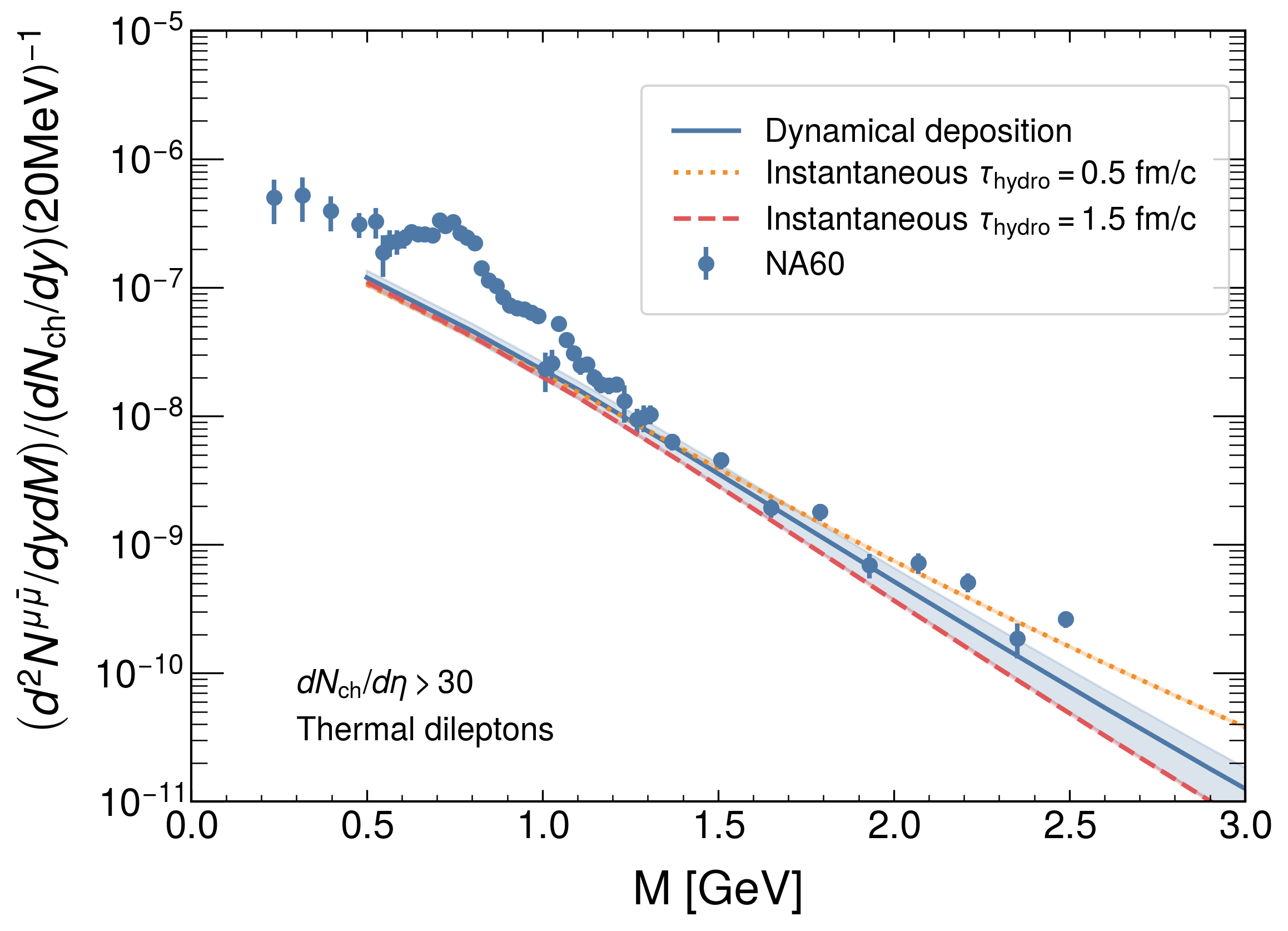}
\includegraphics[width=6cm,clip]{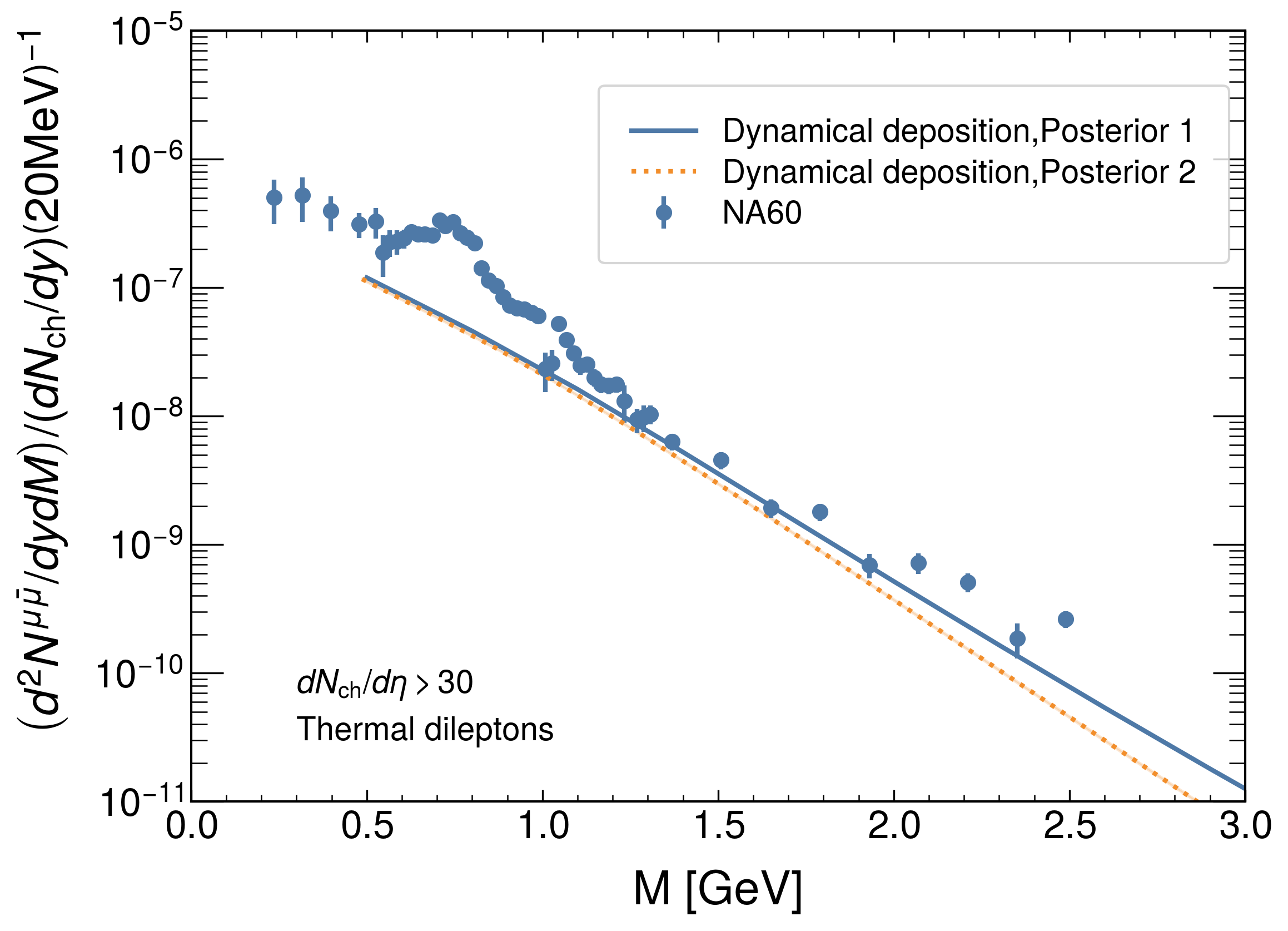}
\caption{Dilepton invariant mass excess spectra (data minus hadronic cocktail) in In+In collisions ($dN_{\rm ch}/d\eta > 30$) from NA60~\cite{NA60:2008dcb}. Left panel: Comparison between dynamical deposition and instantaneous initialization with $\tau_0 = 0.5$ and 1.5 fm/$c$. Right panel: Comparison between different posterior parameter sets. }
\label{fig-2}       % Give a unique label
\end{figure}
In the left panel of Fig.~\ref{fig-2}, the dilepton invariant-mass excess spectra (data minus hadronic cocktail) 
in In+In at 17.3 GeV are shown. We explore the sensitivity of dilepton spectra to the pre-equilibrium stage description in our model. Our default dynamical initial state model continuously deposits energy and momentum to form the QGP medium during the interval $\tau = 0.5 - 1.5$\,fm/$c$ as the nuclei interpenetrate ~\cite{Shen:2023aeg}. The emitted dilepton spectrum is compatible with NA60 data~\cite{NA60:2008dcb}. We further explore the model dependence of dilepton production by studying two extreme scenarios for the pre-equilibrium stage, namely instantaneously depositing all the energy and momentum at a fixed proper time $\tau = 0.5$ fm/$c$, or $\tau = 1.5$ fm/$c$. With the full QGP medium present at $\tau = 0.5$ fm/$c$, more dileptons with large invariant mass are produced at the early time, leading to an even better description of the NA60 data. The scenario with a late-time QGP formation at $\tau = 1.5$\,fm/$c$ limits the space-time volume for dilepton emission in the intermediate mass region (IMR). Our results demonstrate that the IMR of the dilepton spectrum has sensitivity to the early-stage evolution of heavy-ion collisions.

In the right panel of Fig.~\ref{fig-2}, hydrodynamic backgrounds from different posterior parameter sets are used to show their impact on the dilepton excess spectra. Posterior 1~\cite{Jahan:2024wpj} is the default configuration in this proceeding, and Posterior 2 is taken from a recent Bayesian analysis~\cite{Jahan:2025cbp}. Compared with Posterior~1, Posterior~2 includes particle-anti-particle yield ratios in the Bayesian calibration, which shifts the particlization (related to chemical freezeout) to a low switching energy density $e_\mathrm{sw} \sim 0.16$\,GeV/fm$^3$. The two posteriors give similar yields in the low mass region (LMR), while Posterior~1 gives a larger yield in the IMR.
The main reason for this difference is that posterior 2 prefers a larger hotspot size relative to Posterior~1 (0.3 fm vs. 0.1 fm), and the resulting smoother medium is colder during the early time.
This exercise also highlights the potential role played by dileptons for constraining the initial conditions in future Bayesian analyses.

%\section{Conclusion}
To conclude, we employ the Bayesian-calibrated iEBE-MUSIC framework to investigate direct photon and thermal dilepton production at RHIC-BES and CERN SPS energies. 
%We find that thermal photons can reveal early dynamics in low-energy collisions.
An excess of thermal over prompt photons is observed at high $p_T$ in low-energy collisions, suggesting that thermal emission may be measurable.
Thermal dileptons are also sensitive to the pre-equilibrium stage and to the hotspot size in the initial state. 
These results motivate the inclusion of EM probes in future Bayesian analyses to improve our knowledge of QGP properties in the high-temperature regime.

\bigskip
\noindent {\it{Acknowledgments.}}
The work was supported in part by the US National Science Foundation (NSF) under grant number OAC-2004571. It was supported in part by the Natural Sciences and Engineering Research Council of Canada (NSERC) [SAPIN-2020-00048 and SAPIN-2024-00026] (X-Y.W, C.G., and S.J). It was also supported in part by the U.S. Department of Energy, Office of Science, Office of Nuclear Physics, under DOE Award No.~DE-SC0021969 (C.~S.), DE-SC0024232 (C.~S.), DE-SC0024347 (J.-F.P.), and under DOE Contract No.~DE-SC0012704 (B.P.S.) and within the framework of the Saturated Glue (SURGE) Topical Theory Collaboration. C.~S. and J.-F.P. acknowledge DOE Office of Science Early Career Award (DE-SC0021969 and DE-SC0024347). 
Numerical simulations were performed at the Wayne State Grid and the
B\'eluga supercomputer from McGill University, managed by Calcul Qu\'ebec and by the Digital Research Alliance of Canada. 
This research used resources provided by the Open Science Grid (OSG), and supported by the NSF awards \#2030508 and \#1836650. 

%
% BibTeX or Biber users please use (the style is already called in the class, ensure that the "woc.bst" style is in your local directory)
% \bibliography{your_bib_file} % Replace "your_bib_file" with the actual name of your .bib file

\end{document}